\documentclass[11pt,tightenlines,superscriptaddress,floatfixolumn,english,aps,
notitlepage,nofootinbib]{revtex4-1}

\usepackage{amsmath}
\usepackage{amssymb}
\usepackage{esint}
\usepackage{graphicx}
\usepackage{hyperref}
\usepackage{color}
\usepackage[utf8]{inputenc} 
\usepackage{enumitem}
\usepackage{float}
\usepackage{array}
\usepackage{subfig}
\usepackage{soul}
\usepackage[makeroom]{cancel}

\usepackage[dvipsnames]{xcolor}
\usepackage{mathtools, nccmath}
\usepackage{cleveref}  
\usepackage{amsfonts}
\usepackage[normalem]{ulem}

\crefrangeformat{equation}{Eqs. #3(#1)#4--#5(#2)#6}

\newcommand{\mf}{\frac}
\newcommand{\mtf}{\tfrac}
\newcommand{\ml}{\left}
\newcommand{\mr}{\right}

\newcommand{\mfb}{\bar{f}}

\newcommand{\mla}{\langle}
\newcommand{\mra}{\rangle}
\newcommand{\mk}{\kappa_}
\newcommand{\mkg}{\kappa^{(G)}_}
\newcommand{\mkb}{\kappa^{(1,B)}_}
\newcommand{\mbkg}{\bar{\kappa}^{(G)}_}

\begin{document}

\title{
Cumulants from short-range correlations and baryon number conservation - next-to-leading order
}

\author{Micha{\l} Barej}
\email{michal.barej@fis.agh.edu.pl}
\affiliation{AGH University of Science and Technology,
Faculty of Physics and Applied Computer Science,
30-059 Krak\'ow, Poland}

\author{Adam Bzdak}
\email{adam.bzdak@fis.agh.edu.pl}
\affiliation{AGH University of Science and Technology,
Faculty of Physics and Applied Computer Science,
30-059 Krak\'ow, Poland}

\begin{abstract}
We calculate the baryon number cumulants within acceptance with short-range correlations and global baryon number conservation in terms of cumulants in the whole system without baryon conservation. We extract leading and next-to-leading order terms of the large baryon number limit approximation. Our results extend the findings of Refs. \cite{Vovchenko:2020tsr,Barej:2022jij}. These approximations are checked to be very close to the exact results. 
\end{abstract}

\maketitle

\section{Introduction}
The phase diagram of the strongly interacting matter is not yet well explored. Especially, the search for the first-order phase transition between the hadronic matter and quark-gluon plasma, and the corresponding critical endpoint, which are predicted by the effective models, remains a big challenge in high-energy physics \cite{Stephanov:2004wx,BraunMunzinger:2008tz,Braun-Munzinger:2015hba,Bzdak:2019pkr}. It is known that the fluctuations of conserved charges, e.g., baryon number, electric charge, or strangeness are sensitive to the relevant critical phenomena. Therefore, many theoretical projects, as well as experiments in relativistic heavy-ion collisions, have been established to study them \cite{Jeon:2000wg,Asakawa:2000wh,Gazdzicki:2003bb,Gorenstein:2003hk,Stephanov:2004wx,Koch:2005vg,Stephanov:2008qz,Cheng:2008zh,Fu:2009wy,Skokov:2010uh,Stephanov:2011pb,Karsch:2011gg,Schaefer:2011ex,Chen:2011am,Luo:2011rg,Zhou:2012ay,Wang:2012jr,Herold:2016uvv,Luo:2017faz,Szymanski:2019yho,Ratti:2019tvj}.

Cumulants are commonly used to quantify these fluctuations because they naturally appear in statistical mechanics \cite{Stephanov:2008qz,Behera:2018wqk,Acharya:2019izy,Skokov:2011rq,BraunMunzinger:2011ta,Bzdak:2012ab,Braun-Munzinger:2016yjz,Adamczyk:2017wsl,Adare:2015aqk}. On the other hand, the factorial cumulants might be easier to interpret since they represent integrated multiparticle correlation functions  \cite{Botet:2002gj,Ling:2015yau,Bzdak:2016sxg,Bzdak:2019pkr,Adamczyk:2013dal,Bzdak:2016sxg,Bzdak:2018uhv,Bzdak:2018axe,HADES:2020wpc,STAR:2021iop,Vovchenko:2021kxx}. However, both the cumulants and factorial cumulants are affected also by fluctuations unrelated to the phase transition, for instance, the impact parameter fluctuations and the conservation laws, e.g., the baryon number conservation \cite{Skokov:2012ds,Braun-Munzinger:2016yjz,Bzdak:2016jxo,Adamczewski-Musch:2020slf,Kitazawa:2012at,Bzdak:2012an,Braun-Munzinger:2016yjz,Rogly:2018kus,Braun-Munzinger:2019yxj,Acharya:2019izy,Savchuk:2019xfg,Braun-Munzinger:2020jbk,Vovchenko:2021kxx}.

In our previous paper \cite{Barej:2022jij}, we derived analytically the baryon number factorial cumulant generating function in a finite acceptance, assuming short-range correlations and global baryon number conservation. Among other results, we calculated the factorial cumulants and cumulants within the limit of small short-range correlation strengths, $\alpha_k$, and large baryon number, $B$. We also reproduced the relations between cumulants in a subsystem with all correlations and cumulants in the whole system without baryon conservation, initially obtained in Ref. \cite{Vovchenko:2020tsr}.

In this paper, we extend this study and present a method of obtaining the first correction to the cumulants in the large baryon number limit. We also note that the short-range correlations strengths cannot assume arbitrary values. Finally, we compare our approximate analytic results with the brute-force computations.

In the next Section, we show our method of extracting the baryon number cumulants assuming short-range correlations and global baryon number conservation. Then, we present the leading-order and next-to-leading order terms of cumulants in the subsystem with all correlations expanded in the large $B$ limit with respect to cumulants in the whole system without baryon conservation. This is our main result. In the fourth section, we show how our approximate analytic formulas work by comparison with the exact results. The alternative approach and the discussion on the limitations of $\alpha_k$'s originating from the probability theory can be found in the Appendixes.

\section{Method}
\subsection{Previous study}
\begin{figure}[H]
\begin{center}
	\includegraphics[width=0.3\textwidth]{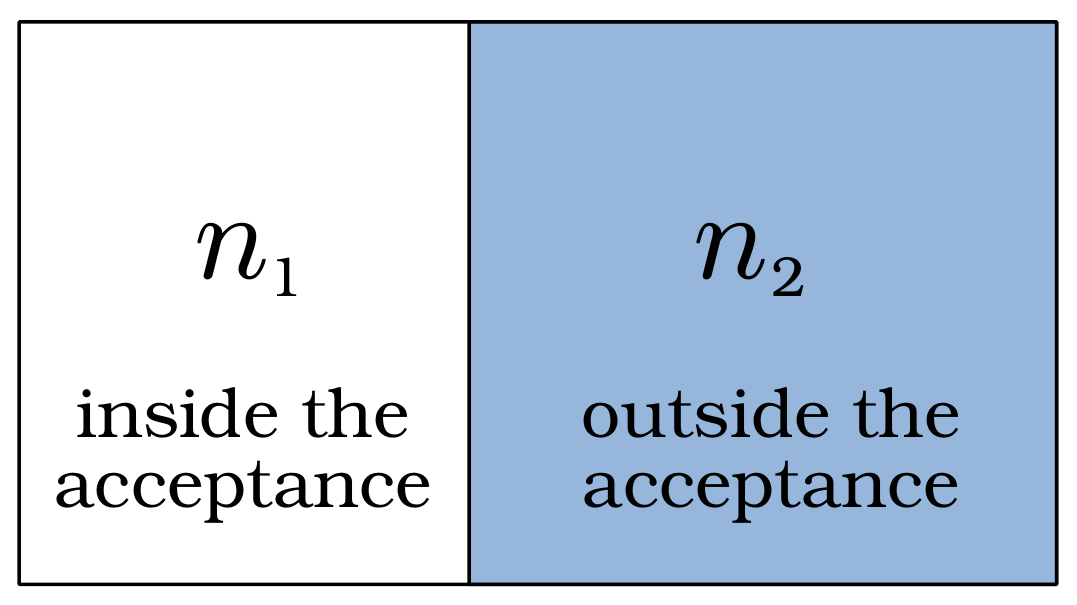}
\caption{The system is divided into two subsystems with $n_1$ particles in the first subsystem (inside the acceptance) and $n_2$ particles in the second one (outside the acceptance). } \label{fig:box}
\end{center}
\end{figure}

In our previous paper \cite{Barej:2022jij}, we considered a system of fixed volume and some number of particles of one kind, say baryons. We divided it into two subsystems (inside and outside the acceptance, see Fig. \ref{fig:box}) which can exchange particles. Let $P_1(n_1)$ and $P_2(n_2)$ be the probabilities that there are $n_1$ particles in the first subsystem and $n_2$ particles in the second one, respectively. The probability that there are $n_1$ particles in the first subsystem and $n_2$ particles in the second one is $P(n_1,n_2) = P_1(n_1)P_2(n_2)$ if there are no correlations between the two subsystems or (approximately) if there are only short-range correlations. Assuming the global baryon number conservation, this probability becomes 
\begin{equation}
P_B(n_1,n_2) = A\;P_1(n_1) P_2(n_2) \delta_{n_1+n_2,B}\,,
\end{equation}
where $A$ is the normalization constant and $B$ is the total baryon number. In this case, the probability that there are $n_1$ particles in the first subsystem (within acceptance) reads 
\begin{equation}
P_{B}(n_1) = \sum_{n_2=0}^{\infty} P_B(n_1,n_2)\,. 
\end{equation}

Then, we calculated the factorial cumulant generating function for the first subsystem with baryon number conservation:
\begin{equation}\label{eq:g_b}
G_{(1,B)}(z) = \ln\left[\frac{A}{B!}\left.\frac{d^B}{dx^B}\exp\left( \sum_{k=1}^{\infty} \frac{(xz-1)^k \hat{C}_k^{(1)} + (x-1)^k \hat{C}_k^{(2)}}{k!} \right) \right|_{x=0} \right]\,,
\end{equation} 
where 
\begin{equation} \label{eq:short-range-fact-cumulants}
\begin{split}
\hat{C}_k^{(1)} &= \langle n_1 \rangle \alpha_k = f \langle N \rangle \alpha_k\,,\\
\hat{C}_k^{(2)}& = \langle n_2 \rangle \alpha_k = (1-f) \langle N \rangle \alpha_k
\end{split}
\end{equation} 
are the short-range factorial cumulants in the first and the second subsystem, respectively (see \cite{Barej:2022jij, Bzdak:2019pkr}), for the multiplicity distribution without global baryon conservation. Here $\langle N \rangle = \langle n_1 \rangle + \langle n_2 \rangle$ is the mean total number of particles in the system, $f = \mla n_1 \mra / \mla N \mra $ is a fraction of particles in the first subsystem, and $\alpha_k$ describes the strength of $k$-particle short-range correlations ($\alpha_1 = 1$). We assumed that the total \textit{average} number of particles $\langle N \rangle = \langle n_1 \rangle + \langle n_2 \rangle = B$. Introducing the global baryon number conservation further requires that the total number of particles $N = n_1 + n_2$ equals $B$ in every event. 

Using the factorial cumulant generating function (\ref{eq:g_b}), one can obtain the factorial cumulants in the first subsystem (within acceptance) with baryon number conservation:
\begin{equation} \label{eq:fact-cum-general}
\hat{C}_k^{(1,B)} = \left.\frac{d^k}{dz^k}G_{(1,B)}(z)\right|_{z=1}\,.
\end{equation}

We obtained \cite{Barej:2022jij} the analytic as well as approximate formulas for the factorial cumulants in the simple case of only two-particle short-range correlations. Then, we also derived the approximate factorial cumulants and cumulants in the limit of $B \to \infty$ assuming multiparticle short-range correlations. These results reproduced the findings of Ref. \cite{Vovchenko:2020tsr} obtained originally using a different approach. 

\subsection{Next-to-leading order correction} \label{subsec:new-method}

We extend the previous study and obtain the next-to-leading order terms of the expansion of the cumulants in the limit of $B \to \infty$. We derive the relevant expressions in two different ways. One method, presented in Appendix \ref{appendix:series}, is based on analyzing the first few terms of the expansions, deducing the following terms, and then summing the infinite series. Here we present another method that is simpler.

In order to make the computations, we approximate Eq. (\ref{eq:g_b}) with Eq. \eqref{eq:short-range-fact-cumulants} by
\begin{equation}\label{eq:g_b-approx}
G_{(1,B)}(z) \approx \ln\left[\frac{A}{B!}\left.\frac{d^B}{dx^B}\left[ \exp\left((xz-1) f B + (x-1) \mfb B \right) \sum_{m=0}^M \mf{V^m}{m!} \right] \right|_{x=0} \right]\,,
\end{equation} 
where $\mfb = 1-f$ and
\begin{equation} \label{eq:V-sum}
V = \sum_{k=2}^{K} \left( \frac{(xz-1)^k}{k!}f B \alpha_k + \frac{(x-1)^k}{k!}\mfb B \alpha_k \right)\,,
\end{equation}
with $M$ and $K$ being the upper limits. Note that in Eq. (\ref{eq:V-sum}) we allow for up to $K$-particle short-range correlations.

To calculate the $B$th derivative with respect to $x$ in Eq. (\ref{eq:g_b-approx}), we use the general Leibnitz formula as described in Ref. \cite{Barej:2022jij}. In this way we evaluate the factorial cumulant generating function (\ref{eq:g_b-approx}) and then the factorial cumulants (\ref{eq:fact-cum-general}). Having the factorial cumulants, we calculate the cumulants according to
\begin{equation} \label{eq:cum-vs-fact-cum}
\kappa_n = \sum_{k=1}^n S(n,k) \hat{C}_k \,,
\end{equation}
where $S(n, k)$ is the Stirling number of the second kind \cite{Friman:2022wuc}.\footnote{See also Appendix A of Ref. \cite{Bzdak:2019pkr} for explicit formulas for the first six cumulants.} For instance, the second cumulant reads
\begin{equation}
\kappa_2^{(1,B)} = \mla n_1 \mra + \hat{C}_2^{(1,B)} \,,
\end{equation}
where $\hat{C}_1^{(1,B)} = \kappa_1^{(1,B)} = \mla n_1 \mra = f B$.

The global (both subsystems combined) short-range factorial cumulants, without the baryon number conservation, are defined as (compare with Eqs. \eqref{eq:short-range-fact-cumulants}):
\begin{equation} \label{eq:fact-cum-glob}
\hat{C}_n^{(G)} = B \alpha_n\,.
\end{equation}
The global cumulants without the baryon number conservation, $\kappa_n^{(G)}$, are obtained by Eq. (\ref{eq:cum-vs-fact-cum}). 

As shown in Ref. \cite{Barej:2022jij}, the cumulants, $\kappa_n^{(1,B)}$, can be expressed as a power series in terms of $B$ where the highest-order term is linear in $B$. Namely,
\begin{equation} \label{eq:cumulant-expansion}
\kappa_n^{(1,B)} \approx \underbrace{ \kappa_n^{(1,B,\text{LO})} }_{u_{n,1} B^1} + \underbrace{ \kappa_n^{(1,B,\text{NLO})} }_{u_{n,0} B^0} + \underbrace{ \kappa_n^{(1,B,\text{NNLO})} }_{u_{n,-1} B^{-1}} + \underbrace{\ldots}_{O(B^{-2})}  \,,
\end{equation}
where $\kappa_n^{(1,B,\text{LO})}$, $\kappa_n^{(1,B,\text{NLO})}$, and $\kappa_n^{(1,B,\text{NNLO})}$ denote the leading-order, next-to-leading-order, and next-to-next-to-leading-order terms of the power series in $B$, respectively.

Let us focus on the second cumulant. Using the method presented in Appendix \ref{appendix:series}, we deduced that in order to extract LO and NLO terms, it is convenient to multiply $\kappa_2^{(1,B)}$ by $(\kappa_2^{(G)})^2 = [B(1 + \alpha_2)]^2$. We define
\begin{equation}
\widetilde{\kappa}_2^{(1,B)} = \kappa_2^{(1,B)} (\kappa_2^{(G)})^2 = \kappa_2^{(1,B)} [B(1 + \alpha_2)]^2 \,.
\end{equation}
Then, we expand $\widetilde{\kappa}_2^{(1,B)}$ into the power series in $\alpha_k$ up to the order of $M$, obtaining $\widetilde{\kappa}_2^{(1,B,\text{ser})}$: 
\begin{equation} \label{eq:k2-series}
\widetilde{\kappa}_2^{(1,B)} \approx \widetilde{\kappa}_2^{(1,B,\text{ser})} =  u_{2,1}(1 + \alpha_2)^2 B^3 +  u_{2,0}(1 + \alpha_2)^2 B^2  + \underbrace{\ldots}_{O(B)}  \,.
\end{equation}
The coefficients of the expansion are calculated as follows,
\begin{equation} \label{eq:u-coeffs}
\begin{split}
{u}_{2,1} &= \mf{1}{(1+ \alpha_2)^2} \lim_{B \to \infty} \mf{\widetilde{\kappa}_2^{(1,B,\text{ser})}}{B^3} \,, \\
{u}_{2,0}  &= \mf{1}{(1+ \alpha_2)^2} \lim_{B \to \infty} \mf{\widetilde{\kappa}_2^{(1,B,\text{ser})} - {u}_{2,1} (1+ \alpha_2)^2 B^3 }{B^2}\,.
\end{split}
\end{equation}
Clearly, it is possible to extract even higher terms in an analogous way. Using Eqs. \eqref{eq:cum-vs-fact-cum} and \eqref{eq:fact-cum-glob}, we express these coefficients in terms of the global short-range cumulants (without the baryon conservation), $\kappa_n^{(G)}$.

The same technique is applied to obtain the leading and next-to-leading order terms of $\kappa_3^{(1,B)}$. Namely, we multiply $\kappa_3^{(1,B)}$ by $(\kappa_2^{(G)})^2$. 

It turns out that $\kappa_4^{(1,B)}$ needs to be multiplied by $(\kappa_2^{(G)})^4$. Namely,
\begin{equation}
\widetilde{\kappa}_4^{(1,B)} = \kappa_4^{(1,B)} (\kappa_2^{(G)})^4 = \kappa_4^{(1,B)} [B(1 + \alpha_2)]^4 \,.
\end{equation}
In this case, equations corresponding to Eqs. \eqref{eq:k2-series} and \eqref{eq:u-coeffs} read:
\begin{equation} 
\widetilde{\kappa}_4^{(1,B)} \approx \widetilde{\kappa}_4^{(1,B,\text{ser})} =  u_{4,1}(1 + \alpha_2)^4 B^5 +  u_{4,0}(1 + \alpha_2)^4 B^4  + \underbrace{\ldots}_{O(B^3)}  \,,
\end{equation}
and
\begin{equation}
\begin{split}
{u}_{4,1} &= \mf{1}{(1+ \alpha_2)^4} \lim_{B \to \infty} \mf{\widetilde{\kappa}_4^{(1,B,\text{ser})}}{B^5} \,, \\
{u}_{4,0}  &= \mf{1}{(1+ \alpha_2)^4} \lim_{B \to \infty} \mf{\widetilde{\kappa}_4^{(1,B,\text{ser})} - {u}_{4,1} (1+ \alpha_2)^4 B^5 }{B^4}\,.
\end{split}
\end{equation}

The results are computed using Mathematica software \cite{mathematica}.

\section{Results}
The relations between cumulants in the first subsystem with all correlations, $\kappa_n^{(1,B)}$, vs. cumulants in the whole system without baryon conservation, $\mkg m$, are given by:
\begin{flalign} \label{eq:k1}
\kappa_1^{(1,B)} &= f B = f \kappa_1^{(G)} \,, &&
\end{flalign}
\begin{flalign} \label{eq:k2-lo}
\kappa_2^{(1,B,\text{LO})} &= \mfb f \kappa_2^{(G)}  \,, &&
\end{flalign}
\begin{flalign} \label{eq:k2-nlo}
\kappa_2^{(1,B,\text{NLO})} &= \mf{1}{2} \mfb f  \mf{ (\mkg3)^2  - \mkg2 \mkg4  }{(\mkg2)^2}  \,, &&
\end{flalign}
\begin{flalign} \label{eq:k3-lo}
\kappa_3^{(1,B,\text{LO})} &= \mfb f (1-2f)\kappa_3^{(G)}  \,, &&
\end{flalign}
\begin{flalign} \label{eq:k3-nlo}
\kappa_3^{(1,B,\text{NLO})} &= \mf{1}{2} f \mfb (1-2f) \mf{ \mkg3 \mkg4  - \mkg2 \mkg5 }{(\mkg2)^2}  \,, &&
\,.
\end{flalign}
\begin{flalign} \label{eq:k4-lo}
\kappa_4^{(1,B,\text{LO})} &=  f \mfb  \ml[ \mkg4 -3f \mfb \ml( \mkg4 + \mf{(\mkg3)^2}{\mkg2} \mr) \mr] \,, &&
\end{flalign}
\begin{flalign} \label{eq:k4-nlo}
\mk4^{(1,B,\text{NLO})} &=\frac{1}{2} f \mfb \ml\{ \frac{\mkg3 \mkg5 - \mkg2 \mkg6}{(\mkg2)^2}  \mr.\\ \nonumber
& +3 f \mfb \ml[ \mf{ 2 (\mkg3)^4 - 5 \mkg2 (\mkg3)^2 \mkg4 + (\mkg2)^2 \mkg3 \mkg5}{(\mkg2)^4}  \ml. + \frac{
   (\mkg4)^2 + \mkg2 \mkg6 }{(\mkg2)^2}   \mr] \mr\}\,. &&
\end{flalign}

The leading-order terms, $\kappa_n^{(1,B,\text{LO})}$, were already obtained in \cite{Barej:2022jij} and originally in \cite{Vovchenko:2020tsr}. The final results for $\kappa_2^{(1,B,NLO)}$ and $\kappa_3^{(1,B,NLO)}$ are obtained already for $M=3$ and $K=5$ (see Eqs. \eqref{eq:g_b-approx} and \eqref{eq:V-sum}). We have checked (up to $M=7$ and $K=7$) that when increasing $M$ or $K$ the results remain unchanged.\footnote{Note that increasing $M$ gives the next terms of $\alpha_k$ power series expansion whereas increasing $K$ allows for higher multiparticle correlations, e.g., $K=6$ allows for up to 6-particle short-range correlations.} The final results for $\kappa_4^{(1,B,NLO)}$ are obtained for $M=5$ and $K=6$. We have verified them up to $M=6$ and $K=7$ as well as up to $M=7$ and $K=6$. When further increasing $M$ or $K$ the computations become challenging.

\section{Examples} \label{sec:examples}
We calculate the cumulants for the selected values of the $k$-particle short-range correlation strengths, $\alpha_k$'s, the fraction of particle number in the first subsystem, $f$, and the baryon number, $B$. We do this in three different ways. Firstly, we calculate the cumulants by a straightforward differentiation using \Crefrange{eq:g_b}{eq:fact-cum-general} (exact results).\footnote{In this method, we calculate up to $B=90$. Calculation of higher derivatives in Eq. \eqref{eq:g_b} becomes challenging.} Secondly, we calculate them using only the leading-order terms (Eqs. (\ref{eq:k2-lo}), (\ref{eq:k3-lo}), and (\ref{eq:k4-lo})). Finally, we calculate them by applying both the leading- and next-to-leading order terms. 

\begin{figure}[H]
\begin{center}
	\includegraphics[width=0.32\textwidth]{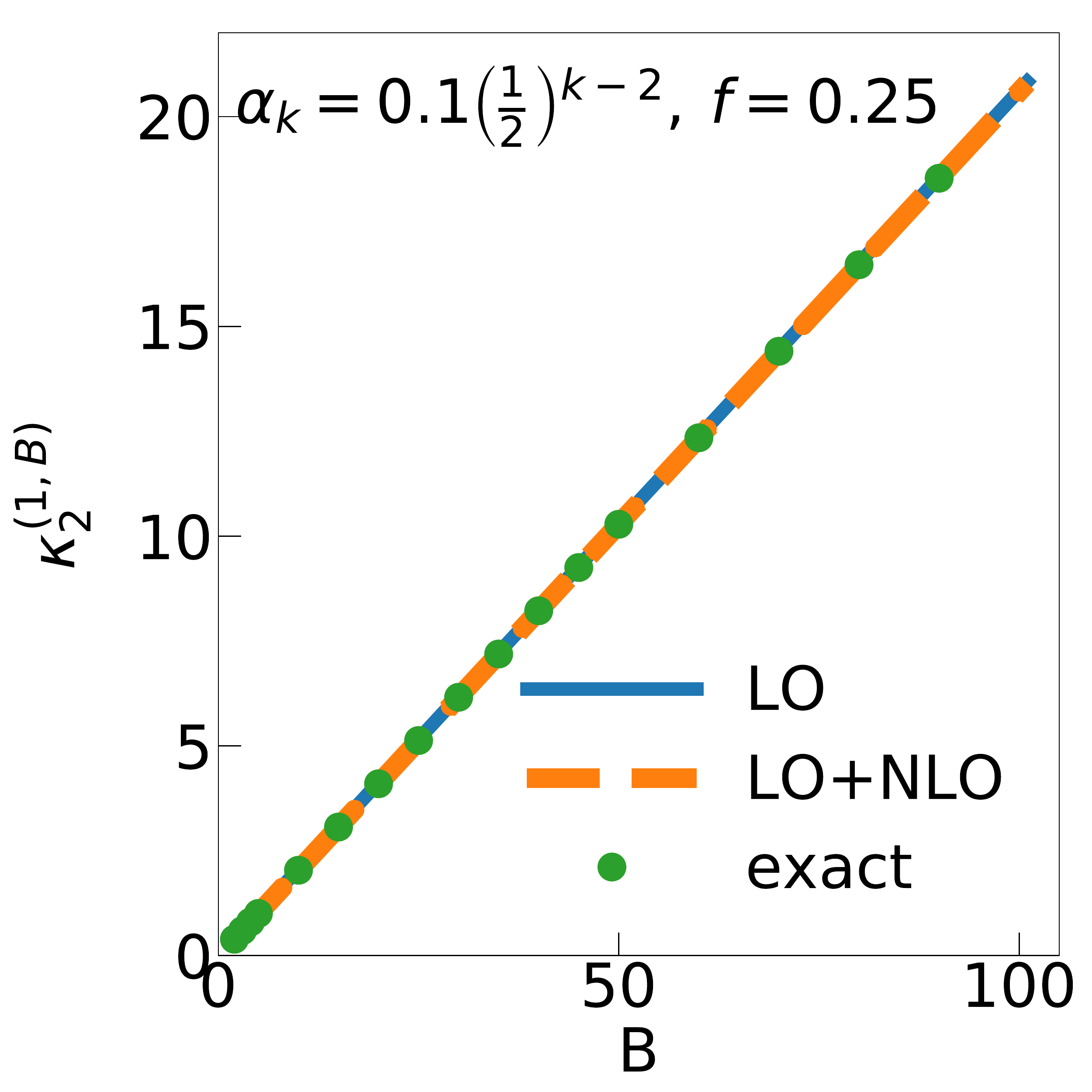}
	\includegraphics[width=0.32\textwidth]{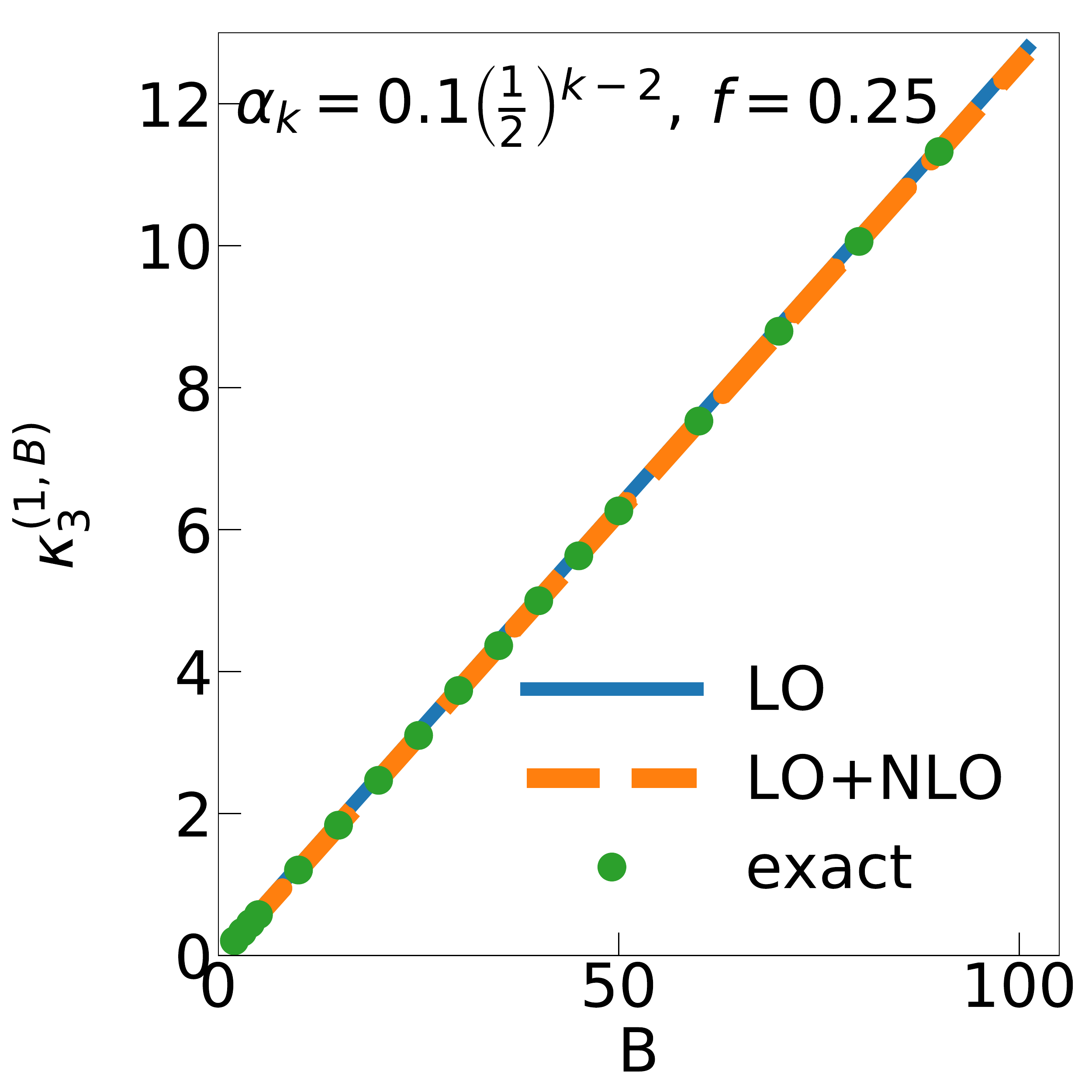} 
	\includegraphics[width=0.32\textwidth]{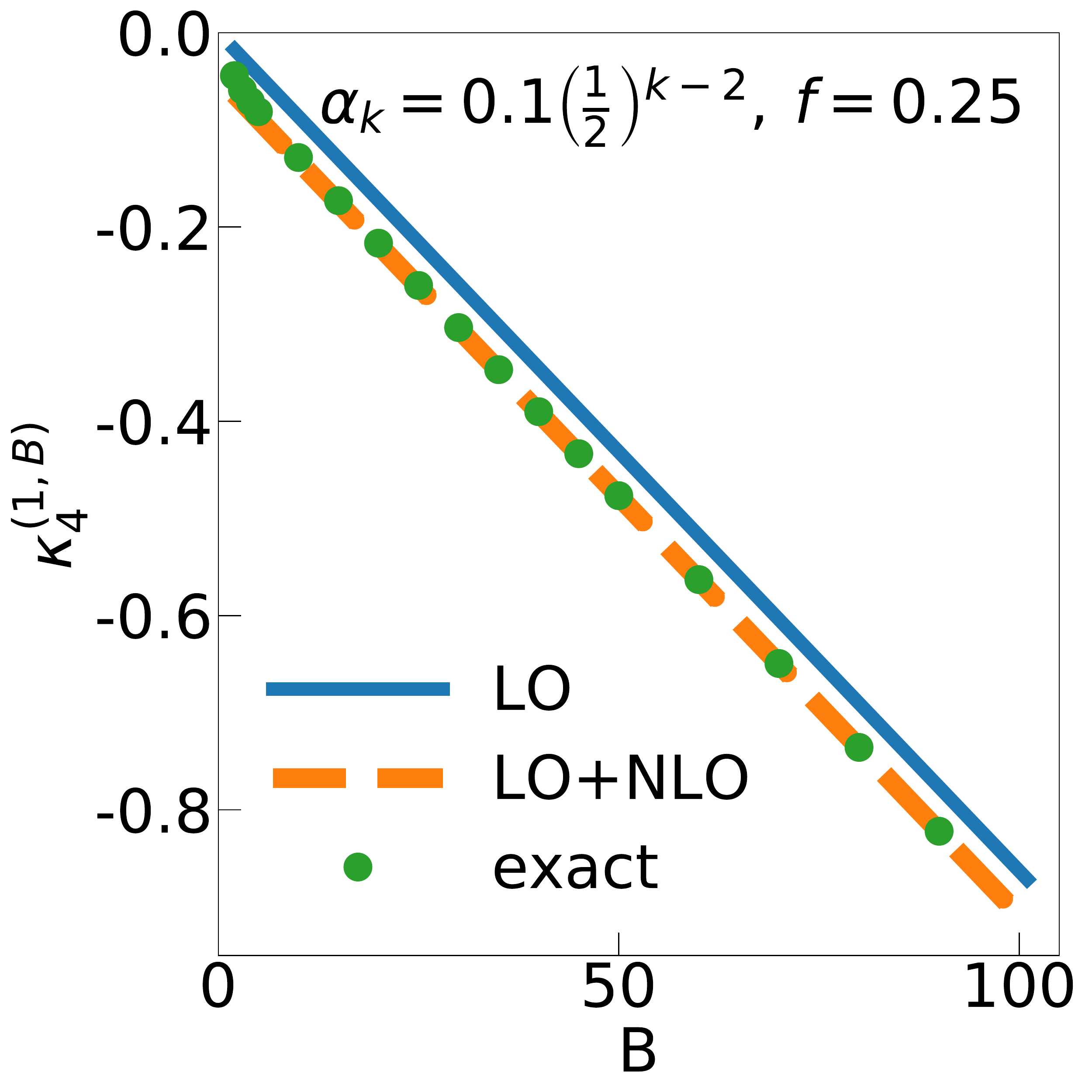}\\
	\includegraphics[width=0.32\textwidth]{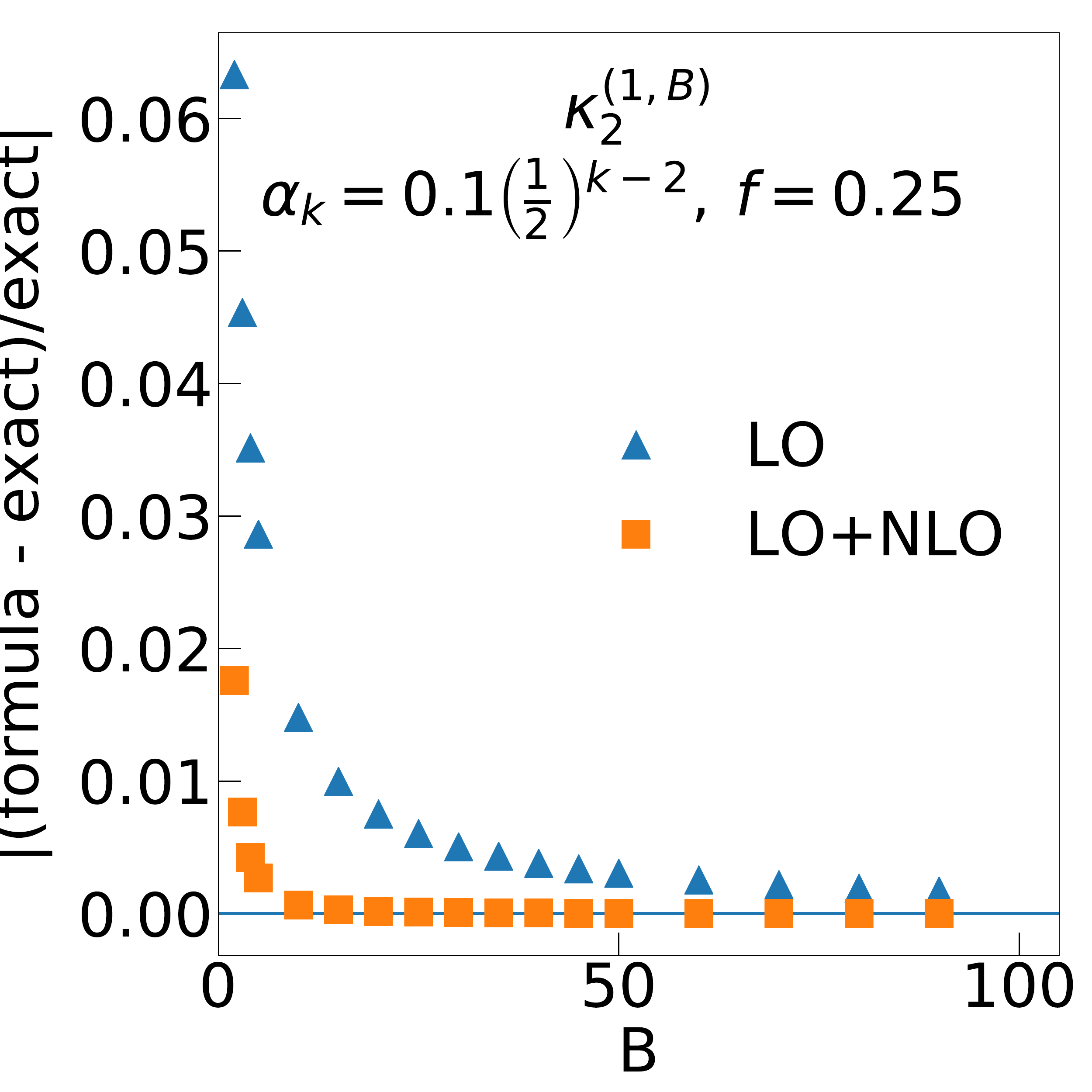}
	\includegraphics[width=0.32\textwidth]{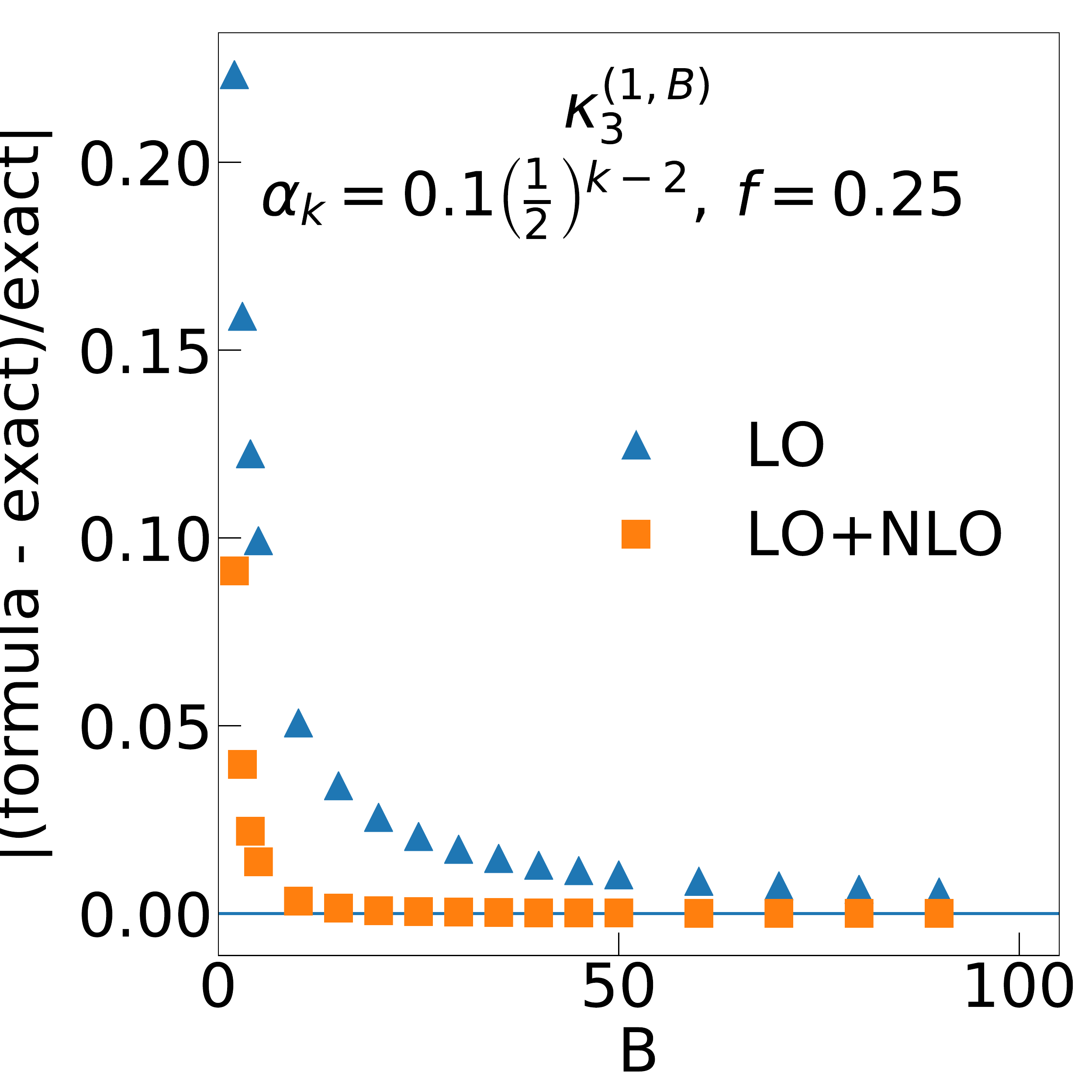}
	\includegraphics[width=0.32\textwidth]{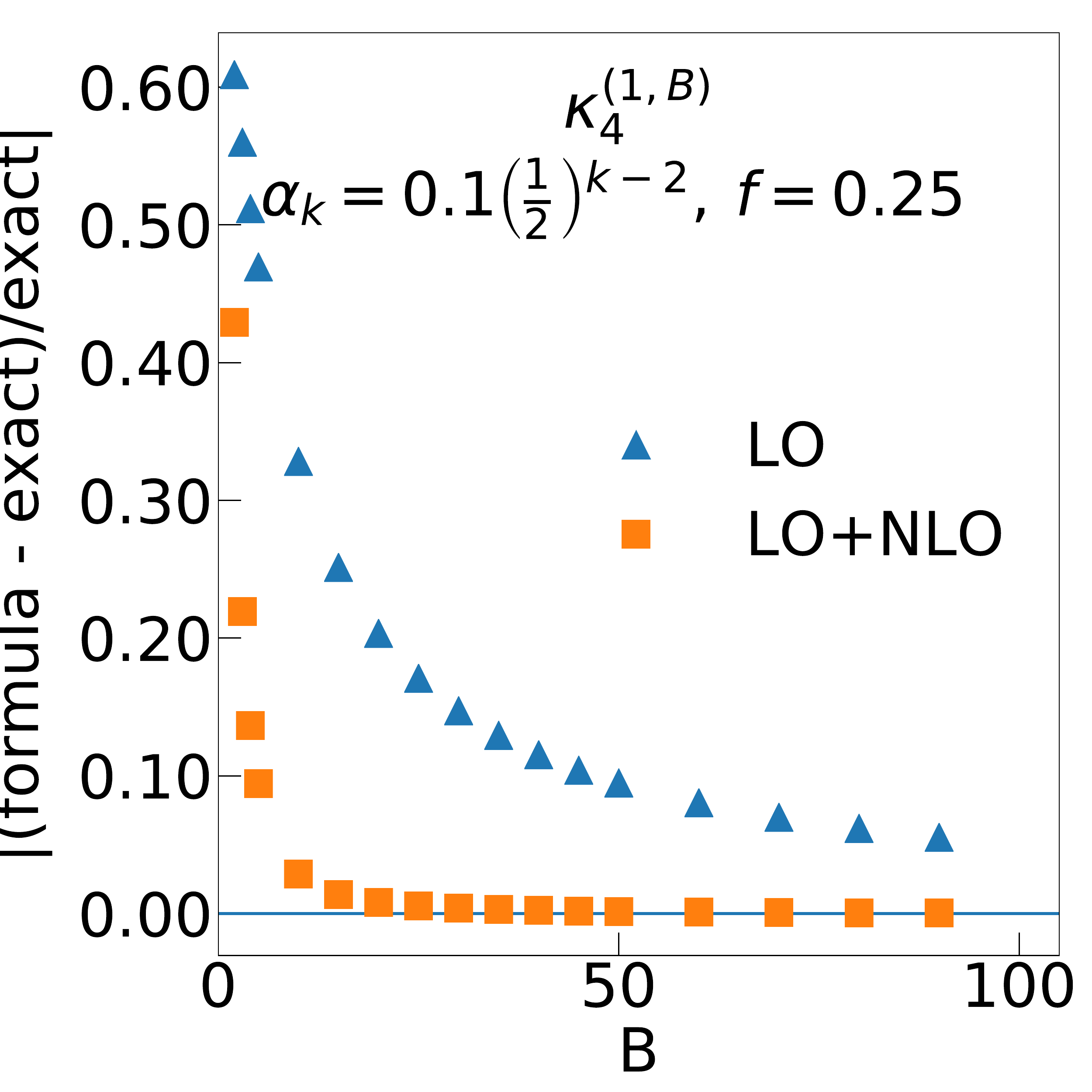}
\caption{Upper plots: The second (left), third (middle), and fourth (right) cumulant in the first subsystem (within acceptance) with the short-range correlations and baryon number conservation as a function of $B$. $\alpha_k = 0.1 \ml(\mf{1}{2}\mr)^{k-2}$, $k=2,3,...,6$, $\alpha_1 = 1$, $f=0.25$. ``LO'' denotes the results obtained from the leading-order terms (Eqs. (\ref{eq:k2-lo}), (\ref{eq:k3-lo}), and (\ref{eq:k4-lo})). ``LO+NLO'' denotes the results obtained using the leading-order and next-to-leading-order terms (Eqs. (\ref{eq:k2-lo}), (\ref{eq:k3-lo}), (\ref{eq:k4-lo}) plus Eqs. (\ref{eq:k2-nlo}), (\ref{eq:k3-nlo}), and (\ref{eq:k4-nlo})). The ``exact'' points denote the direct calculation from \Crefrange{eq:g_b}{eq:fact-cum-general}. The exact results are presented for $B=2,3,4,5,10,15,20,25,30,35,40,45,50,60,70,80,90$. Lower plots: The relative error for each $B$. ``formula'' can be a result of LO or LO+NLO. As seen, the use of the next-to-leading order term improves the results significantly. } \label{fig:vals1-k2-k3-k4}
\end{center}
\end{figure}

The $k$-particle short-range correlations are typically small and the higher-order ones are expected to be smaller than the lower-order ones. Thus, as an example, we study the case of $\alpha_k = 0.1 \ml(\mf{1}{2}\mr)^{k-2}$, $k=2,3,...,6$, $\alpha_1 = 1$, $f=0.25$.\footnote{As seen from \Crefrange{eq:k1}{eq:k4-nlo}, the 6-particle correlation is the highest-order appearing in the LO and NLO terms of the first four cumulants.} In Fig. \ref{fig:vals1-k2-k3-k4}, we show the cumulants as a function of the baryon number, $B$, calculated in three ways and we also present the relative errors with respect to the exact ones. We see that including the NLO term gives a significantly better approximation of the exact results. We have also studied other values of parameters and the observations are consistent. We note that for $f=0.5$, $\mkb3$ vanishes for all values of $B$ \cite{Bzdak:2017ltv,Vovchenko:2020gne}.

It is worth mentioning that $\alpha_k$ cannot be arbitrary. We discuss this issue in Appendix \ref{appendix:limits-alpha}.

\section{Comments and summary}
In this paper, we have extended the results of Ref. \cite{Barej:2022jij}. We have presented the method of obtaining successive terms of the large $B$ limit expansion of the baryon number cumulants in the subsystem with the short-range correlations and global baryon number conservation. We have expressed them by the cumulants in the whole system without baryon number conservation. The newly obtained next-to-leading order terms have improved the approximation as seen in Section \ref{sec:examples}.

In this calculation, we have neglected antibaryons, which makes our results applicable to lower collision energies. We have presented baryon number cumulants, whereas typically in the experiments the proton (net-proton) number cumulants are measured because it is more challenging to detect neutrons.

An interesting but rather challenging extension of this method would be to take into account also antibaryons, as well as other long-range correlations.

\begin{acknowledgements}
This work was partially supported by the Ministry of Science and Higher Education, and by the National Science Centre, Grant No. 2018/30/Q/ST2/00101.
\end{acknowledgements}

\appendix 
\section{The series expansion approach} \label{appendix:series}
Here we present another way of obtaining the leading and next-to-leading order terms of the $B$ power series expansion of the cumulants. 

We obtain the factorial cumulants, $\hat{C}_n^{(1,B)}$, in the same way as in Sec. \ref{subsec:new-method}. Then, we expand them into the $\alpha_k$ power series up to the order of $M$. As shown in \cite{Barej:2022jij}, $\hat{C}_n^{(1,B)}$ can be written as a power series in terms of $B$,
\begin{equation}
\hat{C}_n^{(1,B)} \approx \hat{C}_n^{(1,B,\text{ser})} = \underbrace{ \hat{C}_n^{(1,B,\text{LO})} }_{v_{n,1} B^1} + \underbrace{ \hat{C}_n^{(1,B,\text{NLO})} }_{v_{n,0} B^0} + \underbrace{ \hat{C}_n^{(1,B,\text{NNLO})} }_{v_{n,-1} B^{-1}} + \underbrace{\ldots}_{O(B^{-2})}  \,,
\end{equation}
where $\hat{C}_n^{(1,B,\text{LO})}$, $\hat{C}_n^{(1,B,\text{NLO})}$, and $\hat{C}_n^{(1,B,\text{NNLO})}$ denote the leading, next-to-leading, and next-to-next-to-leading order terms. The coefficients, $v_{n,1}$ and $v_{n,0}$, are calculated as follows:
\begin{equation}
\begin{split}
v_{n,1} &= \lim_{B \to \infty} \mf{\hat{C}_n^{(1,B,\text{ser})}}{B} \,, \\
v_{n,0}  &= \lim_{B \to \infty} \ml(\hat{C}_n^{(1,B,\text{ser})} - v_{n,1} B \mr)\,.
\end{split}
\end{equation}
The subsequent coefficients can be obtained in a similar way. Note that the leading-order terms, $\hat{C}_n^{(1,B,\text{LO})}$, were presented in Ref. \cite{Barej:2022jij}. 

We calculate the LO and NLO terms of cumulants from factorial cumulants using Eq. \eqref{eq:cum-vs-fact-cum}. For instance, 
\begin{equation}
\begin{split}
\kappa_3^{(1,B)} &= \mla n_1 \mra + 3 \hat{C}_2^{(1,B)} + \hat{C}_3^{(1,B)} \\  
&\approx f B + 3\ml( \hat{C}_2^{(1,B,\text{LO})} + \hat{C}_2^{(1,B,\text{NLO})} \mr) + \ml( \hat{C}_3^{(1,B,\text{LO})} + \hat{C}_3^{(1,B,\text{NLO})} \mr) \\
&= \underbrace{\ml(fB + 3\hat{C}_2^{(1,B,\text{LO})} + \hat{C}_3^{(1,B,\text{LO})} \mr)}_{\kappa_3^{(1,B,\text{LO})}} + \underbrace{ \ml( 3\hat{C}_2^{(1,B,\text{NLO})} + \hat{C}_3^{(1,B,\text{NLO})} \mr) }_{ \kappa_3^{(1,B,\text{NLO})} } \,,
\end{split}
\end{equation}
where $\hat{C}_1^{(1,B)} = \kappa_1^{(1,B)} = \mla n_1 \mra = f B$. 

Next, we express $\kappa_n^{(1,B,\text{LO})}$ and $\kappa_n^{(1,B,\text{NLO})}$ by the global (in the whole system) factorial cumulants, $\hat{C}_n^{(G)}$, with short-range correlations but without baryon number conservation, defined in Eq. \eqref{eq:fact-cum-glob}, and then, by the global cumulants without the baryon number conservation, $\kappa_n^{(G)}$. 

For the Poisson distribution, all the cumulants $\kappa_n^{(G)} = \mla N \mra = B$. Therefore, 
\begin{equation} \label{eq:kappa-bar}
\bar{\kappa}_n^{(G)} = \mf{\kappa_n^{(G)}}{B} - 1 
\end{equation}
describes the deviation from the Poisson distribution. We expect it to be small (for example, $\bar{\kappa}_2^{(G)} = \alpha_2$) and thus we express $\kappa_n^{(1,B,\text{LO})}$ and $\kappa_n^{(1,B,\text{NLO})}$ in terms of $\bar{\kappa}_m^{(G)}$.

We note that when using Eqs. \eqref{eq:g_b-approx} and \eqref{eq:V-sum}, the final results for $\kappa_1^{(1,B)}$, $\kappa_2^{(1,B,\text{LO})}$, and $\kappa_3^{(1,B,\text{LO})}$ are obtained already for $M=1$ and $K=3$, and an increase of $M$ and $K$ does not modify the results. In other cases, we obtain more and more terms when increasing $M$. However, we can deduce empirically the series in $\mbkg2$ using the first few terms and confirm them by increasing $M$ and $K$ in Eqs. (\ref{eq:g_b-approx}) and (\ref{eq:V-sum}). We have performed the computations up to $M = 8$, $K=6$, and also up to $M=4$, $K=8$. By summing the series up to infinity, we derive the finite analytic formulas for LO and NLO terms of the cumulants, and we obtain the same results as given by \Crefrange{eq:k1}{eq:k4-nlo}.

\subsection{$\kappa_2^{(1,B)}$ and $\kappa_3^{(1,B)}$}
The leading-order terms of $\kappa_2^{(1,B)}$ and $\kappa_3^{(1,B)}$ read \cite{Barej:2022jij,Vovchenko:2020tsr}:
\begin{equation} 
\begin{split}
\kappa_2^{(1,B,\text{LO})} &= \mfb f \kappa_2^{(G)}  \,, \\
\kappa_3^{(1,B,\text{LO})} &= \mfb f (1-2f)\kappa_3^{(G)}  \,.
\end{split}
\end{equation}

The NLO term of $\kappa_2^{(1,B)}$ is given by the power series in $\mbkg2$. This is understandable since in the process of computation we expand the factorial cumulants into power series about $\alpha_k=0$ up to order $M$ (see Sec. \ref{subsec:new-method}). The NLO term of the second cumulant with $M=8$ is given by
\begin{equation}
\begin{split}
\kappa_2^{(1,B,\text{NLO})} &\approx \mf{1}{2} f \mfb \biggl[ -\mbkg2 + 2 (\mbkg2)^2 - 3 (\mbkg2)^3 + 4 (\mbkg2)^4 - 5 (\mbkg2)^5 + 6 (\mbkg2)^6 -7 (\mbkg2)^7 + 8 (\mbkg2)^8  \\
& \phantom{=\mf{1}{2} f } +2\mbkg3 \ml( 1 - 2 \mbkg2 + 3 (\mbkg2)^2 - 4 (\mbkg2)^3 + 5 (\mbkg2)^4 - 6 (\mbkg2)^5 + 7 (\mbkg2)^6 - 8 (\mbkg2)^7 \mr) \\
& \phantom{=\mf{1}{2} f } +(\mbkg3)^2 \ml( 1 - 2 \mbkg2 + 3 (\mbkg2)^2 - 4 (\mbkg2)^3 + 5 (\mbkg2)^4 - 6 (\mbkg2)^5 + 7 (\mbkg2)^6 \mr) \\
& \phantom{=\mf{1}{2} f } +\mbkg4 \ml(
-1 + \mbkg2 - (\mbkg2)^2 + (\mbkg2)^3 - (\mbkg2)^4 + (\mbkg2)^5 - (\mbkg2)^6 + (\mbkg2)^7 \mr) \biggr]\,.
\end{split}
\end{equation}
Here we recognize the series:
\begin{equation}
\begin{split}
\kappa_2^{(1,B,\text{NLO})} &\approx \mf{1}{2} f \mfb \biggl[\sum_{n=1}^{N=8} n \ml(-\mbkg2 \mr)^n + 2\mbkg3 \sum_{n=1}^{N=8} n \ml(-\mbkg2 \mr)^{n-1} + (\mbkg3)^2 \sum_{n=1}^{N=7} n \ml(-\mbkg2 \mr)^{n-1} \\
& \phantom{ = \mf{1}{2} f \mfb} - \mbkg4 \sum_{n=1}^{N=8} \ml(-\mbkg2 \mr)^{n-1}    \biggr]\,.
\end{split}
\end{equation}
Then, we assume that the next terms follow this pattern and we sum up with $N \to \infty$. We obtain
\begin{equation} 
\begin{split}
\kappa_2^{(1,B,\text{NLO})} &= \mf{1}{2} \mfb f  \biggl[-\mf{\mbkg2}{(1+\mbkg2)^2} + \mf{2\mbkg3}{(1+\mbkg2)^2} + \mf{(\mbkg3)^2}{(1+\mbkg2)^2} -\mf{\mbkg4}{1+\mbkg2}   \biggr] \\
&=-\mf{1}{2} \mfb f  \mf{ \mbkg2 - 2\mbkg3  + \mbkg4(1 + \mbkg2) - (\mbkg3)^2  }{(1+\mbkg2)^2}
\,.
\end{split}
\end{equation}
We substitute back Eq. \eqref{eq:kappa-bar} and we obtain Eq. \eqref{eq:k2-nlo}.

The case of the NLO term of $\kappa_3^{(1,B)}$ is similar. The result reads:
\begin{equation}
\begin{split}
\kappa_3^{(1,B,\text{NLO})} &= \mf{1}{2} f \mfb (1-2f) \biggl[\sum_{n=1}^{N=8} n \ml(-\mbkg2 \mr)^n + \mbkg3 \sum_{n=1}^{N=8} n \ml(-\mbkg2 \mr)^{n-1} + \mbkg4 \sum_{n=1}^{N=8} n \ml(-\mbkg2 \mr)^{n-1} \\
& \phantom{ = \mf{1}{2} f \mfb(1-2f)} + \mbkg3 \mbkg4 \sum_{n=1}^{N=7} n\ml(-\mbkg2 \mr)^{n-1} -\mbkg5 \sum_{n=1}^{N=8} \ml(-\mbkg2 \mr)^{n-1}   \biggr]\,.
\end{split}
\end{equation}
By letting $N \to \infty$, we have
\begin{equation} 
\begin{split}
\kappa_3^{(1,B,\text{NLO})} &= \mf{1}{2} f \mfb (1-2f)  \biggl[-\mf{\mbkg2}{(1+\mbkg2)^2} + \mf{\mbkg3}{(1+\mbkg2)^2} + \mf{\mbkg4}{(1+\mbkg2)^2} + \mf{\mbkg3 \mbkg4}{(1+\mbkg2)^2} - \mf{\mbkg5}{1+\mbkg2}    \biggr] \\
&= -\mf{1}{2} f \mfb (1-2f) \mf{ \mbkg2 - \mbkg3 - \mbkg4  + \mbkg5(1 + \mbkg2) - \mbkg3 \mbkg4}{(1+\mbkg2)^2} \,.
\end{split}
\end{equation}
Using Eq. \eqref{eq:kappa-bar}, we obtain Eq. \eqref{eq:k3-nlo}.

\subsection{$\kappa_4^{(1,B)}$}
\subsubsection{Leading-order term}
In the case of $\kappa_4^{(1,B)}$, the series appear already in the LO term. Here we obtain:
\begin{equation}
\begin{split}
\kappa_4^{(1,B,\text{LO})} &\approx  f \mfb B \biggl[1 - 6f \mfb + (1-3f \mfb) \mbkg4\\
&\phantom{= f \mfb \biggl[} - 3f \mfb \biggl(\sum_{n=1}^{N=8} \ml(-\mbkg2 \mr)^n + 2\mbkg3 \sum_{n=1}^{N=8} \ml(-\mbkg2 \mr)^{n-1} + (\mbkg3)^2 \sum_{n=1}^{N=7} \ml(-\mbkg2 \mr)^{n-1} \biggr) \biggr] \,.
\end{split}
\end{equation}
After applying $N \to \infty$:
\begin{equation}
\begin{split}
\kappa_4^{(1,B,\text{LO})} &=  f \mfb B \biggl[1 - 6f \mfb + (1-3f \mfb) \mbkg4 - 3f \mfb \biggl( -\mf{\mbkg2}{1+\mbkg2} + \mf{2\mbkg3}{1+\mbkg2} + \mf{(\mbkg3)^2}{1+\mbkg2} \biggr) \biggr] \\
&= f \mfb B \biggl[1 - 6f \mfb + (1-3f \mfb) \mbkg4 - 3f \mfb \mf{(\mbkg3)^2 + 2\mbkg3 - \mbkg2}{1 + \mbkg2} \biggr] \,.
\end{split}
\end{equation}

Then, using Eq. \eqref{eq:kappa-bar}, we obtain Eq. \eqref{eq:k4-lo} which is exactly in agreement with the results of Refs. \cite{Barej:2022jij,Vovchenko:2020tsr}. This confirms that our method of series expansion works.

\subsubsection{Next-to-leading order term}
Now we focus on the first correction (next-to-leading-order) term of $\kappa_4^{(1,B)}$. Here, we obtained a much more complicated series:
\vfill
\begin{subequations} \label{eq:k4nlo-sers}
\begin{align} 
\mf{\mk4^{(1,B,\text{NLO})}}{-\mf{1}{2} f \mfb} =& \ml(\mbkg2 -2(\mbkg2)^2 + 3(\mbkg2)^3 - 4 (\mbkg2)^4 + 5 (\mbkg2)^5 - 6 (\mbkg2)^6 + 7 (\mbkg2)^7 \mr. \nonumber \\
& \ml. - 8 (\mbkg2)^8 \mr)\\
& + f \mfb \ml(-6\mbkg2 + 9(\mbkg2)^2 -3(\mbkg2)^3 -18(\mbkg2)^4 + 60(\mbkg2)^5 -129(\mbkg2)^6 \mr. \nonumber \\
& \phantom{+ ff (} \ml. + 231(\mbkg2)^7 -372(\mbkg2)^8 \mr) \\
& + \mbkg3 \ml(-1 +2\mbkg2 -3(\mbkg2)^2 +4(\mbkg2)^3 -5(\mbkg2)^4 +6(\mbkg2)^5 -7(\mbkg2)^6 \mr. \nonumber \\
& \phantom{+ \mbkg3 ( } \ml. +8(\mbkg2)^7 \mr) \\
& + f \mfb \mbkg3 \ml(3 +12\mbkg2 -69 (\mbkg2)^2 + 192(\mbkg2)^3 -405(\mbkg2)^4 +732(\mbkg2)^5 \mr. \nonumber \\
& \phantom{= f f \mbkg3 1} \ml. -1197(\mbkg2)^6 +1824(\mbkg2)^7 \mr) \\
& + f \mfb (\mbkg3)^2 \ml(-21 + 99\mbkg2 -270(\mbkg2)^2 + 570(\mbkg2)^3 -1035(\mbkg2)^4 \mr. \nonumber \\
& \phantom{=  f f \mbkg3 123} \ml. + 1701(\mbkg2)^5 -2604(\mbkg2)^6 \mr) \\
& + f \mfb (\mbkg3)^3 \ml(-24 +96\mbkg2 -240(\mbkg2)^2 + 480(\mbkg2)^3 -840(\mbkg2)^4 \mr. \nonumber \\ 
& \phantom{= f f \mbkg3 123} \ml. + 1344(\mbkg2)^5 \mr) \\
& + f \mfb (\mbkg3)^4 \ml(-6 +24\mbkg2 -60(\mbkg2)^2 +120(\mbkg2)^3 - 210(\mbkg2)^4 \mr) \\
& + f \mfb \mbkg4 \ml(9 -33\mbkg2 +72 (\mbkg2)^2 - 126(\mbkg2)^3 +195(\mbkg2)^4 -279(\mbkg2)^5 \mr. \nonumber \\
& \phantom{= \mbkg3 123} \ml. +378(\mbkg2)^6 -492(\mbkg2)^7 \mr) \\
& + f \mfb \mbkg3 \mbkg4 \ml(30 -90\mbkg2 + 180(\mbkg2)^2 - 300(\mbkg2)^3 + 450(\mbkg2)^4 \mr. \nonumber \\
& \phantom{= \mbkg3 \mbkg4 123} \ml. -630(\mbkg2)^5 + 840(\mbkg2)^6 \mr) \\
& + f \mfb (\mbkg3)^2 \mbkg4 \ml(15 -45\mbkg2 + 90(\mbkg2)^2 - 150(\mbkg2)^3 +225(\mbkg2)^4 \mr. \nonumber \\
& \phantom{= (\mbkg3)^2 \mbkg4 123} \ml. - 315(\mbkg2)^5 \mr) \\
& + f \mfb (\mbkg4)^2 \ml(-3 +6\mbkg2 -9(\mbkg2)^2 +12(\mbkg2)^3 -15(\mbkg2)^4 +18(\mbkg2)^5 \mr. \nonumber \\
& \phantom{= (\mbkg3)^2 123} \ml. - 21(\mbkg2)^6 \mr) \\
& + (1+ 3f \mfb) \mbkg5 \ml(-1 +2\mbkg2 -3(\mbkg2)^2 +4(\mbkg2)^3 -5(\mbkg2)^4 +6(\mbkg2)^5 \mr. \nonumber \\
& \phantom{= (1+ 3f \mfb) \mbkg5 } \ml. -7(\mbkg2)^6 + 8(\mbkg2)^7 \mr) \\
& + (1+ 3f \mfb) \mbkg3 \mbkg5 \ml(-1 +2\mbkg2 -3(\mbkg2)^2 +4(\mbkg2)^3 -5(\mbkg2)^4 \mr. \nonumber \\
& \phantom{= (1+ 3f \mfb) \mbkg3 \mbkg5 } \ml. +6(\mbkg2)^5 - 7(\mbkg2)^6 \mr) \\
& + (1- 3f \mfb) \mbkg6 \ml(1 -\mbkg2 +(\mbkg2)^2 -(\mbkg2)^3 +(\mbkg2)^4 - (\mbkg2)^5 \mr. \nonumber \\
& \phantom{= (1+ 3f \mfb) \mbkg3 \mbkg5  } \ml. +(\mbkg2)^6 - (\mbkg2)^7 \mr) \,,
\end{align}
\end{subequations}
where for readability we divide $\mk4^{(1,B,\text{NLO})}$ by $(-\mf{1}{2} f \mfb)$.

Now we address the series one by one, assuming, as before that the remaining terms of the series (up to infinity) follow the same patterns. For reference, we number the series in Eq. (\ref{eq:k4nlo-sers}) by the line letters.

\paragraph{Series a, c, k, l, m, n. }Some of the series are easy to calculate:
\begin{equation}
\text{a}: \quad -\sum_{n=1}^\infty n (-\mbkg2)^n = \mf{\mbkg2}{(1+ \mbkg2)^2}\,,
\end{equation}
\begin{equation}
\text{c}: \quad -\sum_{n=1}^\infty n (-\mbkg2)^{n-1} = -\mf{1}{(1+ \mbkg2)^2}\,,
\end{equation}
\begin{equation}
\text{n}: \quad \sum_{n=0}^\infty (-\mbkg2)^{n} = \mf{1}{1+ \mbkg2}\,.
\end{equation}
l and m are the same as c. k is just 3 times c.

The other series are less obvious.

\paragraph{Series i and j. } We begin with i and denote the coefficient by $a_n$.
\begin{equation}
\begin{split}
\text{i}: &\quad 30\ml(1 -3\mbkg2 + 6(\mbkg2)^2 - 10(\mbkg2)^3 + 15(\mbkg2)^4 - 21(\mbkg2)^5 + 28(\mbkg2)^6 + \ldots \mr) = \\ 
&= 30 \sum_{n=1}^{\infty} a_n (-\mbkg2)^{n-1}  \,.
\end{split}
\end{equation}
Note that $a_n = \sum_{i=1}^{n} i = \mf{n(n+1)}{2} $ - each $a_n$ is a sum of the arithmetic sequence.\\
Therefore,
\begin{equation}
\text{i}: \quad 30 \sum_{n=1}^{\infty} \mf{n(n+1)}{2}  (-\mbkg2)^{n-1} = \mf{30}{(1 + \mbkg2)^3}   \,.
\end{equation}
Similarly in j:
\begin{equation}
\text{j}: \quad 15 \sum_{n=1}^{\infty} \mf{n(n+1)}{2}  (-\mbkg2)^{n-1} = \mf{15}{(1 + \mbkg2)^3}   \,.
\end{equation}

\paragraph{Series h. }For h the situation is similar:
\begin{equation}
\begin{split}
\text{h}: &\quad 3 \ml(3 -11\mbkg2 + 24(\mbkg2)^2 - 42(\mbkg2)^3 + 65(\mbkg2)^4 - 93(\mbkg2)^5 + 126(\mbkg2)^6 \mr. \\
& \phantom{3 3 333} \ml. - 164(\mbkg2)^7 + \ldots \mr) = 3 \sum_{n=1}^{\infty} a_n (-\mbkg2)^{n-1}  \,,
\end{split}
\end{equation}
where we deduce how to obtain the subsequent terms by observing that:\\
$a_1 = 3$, \\
$a_2 = 3 + (3 + 5 \cdot 1) = 11$, \\
$a_3 = 3 + (3 + 5 \cdot 1) + (3 + 5 \cdot 2) = 24$, \\
...\\
$a_n = 3 + (3 + 5 \cdot 1) + (3 + 5 \cdot 2) + ... + (3 + 5(n-1))$. \\
$a_n$ is a sum of $n$ terms which we call $b_m$: $a_n =\sum_{m=1}^n b_m $, where $b_m = 3 + 5(m-1)$. Thus,
\begin{equation}
a_n =\sum_{m=1}^n b_m = \sum_{m=1}^n [3 + 5(m-1)] = \mf{(5n+1)n}{2}
\end{equation}
\begin{equation}
\text{h}: \quad 3 \sum_{n=1}^{\infty} \mf{n(5n+1)}{2}  (-\mbkg2)^{n-1} = -\mf{3(2\mbkg2 - 3)}{(1 + \mbkg2)^3}   \,.
\end{equation}

\paragraph{Series f and g. } The series in line f has one more level of complexity:
\begin{equation}
\begin{split}
\text{f}: &\quad -24 \ml(1 -4\mbkg2 +10(\mbkg2)^2 -20(\mbkg2)^3 +35(\mbkg2)^4 -56(\mbkg2)^5 \mr) = \\
&= -24 \sum_{n=1}^{\infty} a_n (-\mbkg2)^{n-1}  \,,
\end{split}
\end{equation}
where:\\
$a_1 = 1$,\\
$a_2 = 1 + (1 + 2) = 4$,\\
$a_3 = 1 + (1 + 2) + (1 + 2 + 3) = 10$,\\
$a_4 = 1 + (1 + 2) + (1 + 2 + 3) + (1 + 2 + 3 + 4) = 20$,\\
...\\
$a_n = 1 + (1 + 2) + ... + (1 + 2 + ... + n)$.\\
Again, $a_n = \sum_{m=1}^n b_m$,
where $b_m  = 1 + 2 + ... + m = \mf{m(m+1)}{2}$. Then,
\begin{equation}
a_n =  \sum_{m=1}^{n} b_m = \sum_{m=1}^{n} \mf{m(m+1)}{2} = \mf{n(n+1)(n+2)}{6}\,.
\end{equation}
Eventually,
\begin{equation}
\text{f}: \quad -24 \sum_{n=1}^{\infty} \mf{n(n+1)(n+2)}{6}  (-\mbkg2)^{n-1} = -\mf{24}{(1 + \mbkg2)^4}   \,.
\end{equation}
The same series appears in g:
\begin{equation}
\text{g}: \quad -6 \sum_{n=1}^{\infty} \mf{n(n+1)(n+2)}{6}  (-\mbkg2)^{n-1} = -\mf{6}{(1 + \mbkg2)^4}   \,.
\end{equation}

\paragraph{Series e. } e is similar:
\begin{equation}
\begin{split}
\text{e}: &\quad -3 \ml(7 -33\mbkg2 +90(\mbkg2)^2 -190(\mbkg2)^3 + 345(\mbkg2)^4 -567(\mbkg2)^5 + 868(\mbkg2)^6  \mr) = \\
&= -3 \sum_{n=1}^{\infty} a_n (-\mbkg2)^{n-1}  \,,
\end{split}
\end{equation}
where\\
$a_1 = 7$,\\
$a_2 = 7 + [7 + (7 + 12)] = 33$,\\
$a_3 = 7 + [7 + (7 + 12)] + [7 + (7 +12) + (7+ 2\cdot 12)] = 90$,\\
$a_4 = 7 + [7 + (7 + 12)] + [7 + (7 +12) + (7+ 2\cdot 12)] + [7 + (7 +12) + (7+ 2\cdot 12) + (7+ 3\cdot 12)] = 190$,\\
$a_n = 7 + [7 + (7 + 12)] + ... + [7 + (7 +12) + (7+ 2\cdot 12) + ... + (7+ (n-1)\cdot 12)]$.\\
We denote it as: $a_n = \sum_{m=1}^n b_m$, where: $b_m = \sum_{k=1}^{m}[7 + (k-1)12] = (6m+1)m$. Thus,
\begin{equation}
a_n  = \sum_{m=1}^{n}b_m = \sum_{m=1}^{n} (6m+1)m = \mf{n(n+1)(4n+3)}{2}
\end{equation}

\begin{equation}
\text{e}: \quad -3 \sum_{n=1}^{\infty} \mf{n(n+1)(4n+3)}{2}  (-\mbkg2)^{n-1} = \mf{3(5\mbkg2 -7)}{(1 + \mbkg2)^4}   \,.
\end{equation}

\paragraph{Series d.}
Now we focus on d:
\begin{equation}
\begin{split}
\text{d}: \quad &-1(-3) +2(6\mbkg2) -3(23(\mbkg2)^2) +4(48(\mbkg2)^3) -5(81(\mbkg2)^4) + 6(122(\mbkg2)^5)\\
& -7(171(\mbkg2)^6) + 8(228(\mbkg2)^7) = -\sum_{n=1}^{\infty}n a_n (-\mbkg2)^{n-1}  \,,
\end{split}
\end{equation}
where:\\
$a_1 = -3$,\\
$a_2 = -3 + 9 = 6$,\\
$a_3 = -3 + 9 + (9 + 8) = 23$,\\
$a_4 = -3 + 9 + (9 + 8) + (9 + 2\cdot 8) = 48$,\\
$a_5 = -3 + 9 + (9 + 8) + (9 + 2\cdot 8) + (9 + 3\cdot 8) = 81$,\\
...\\
$a_n = -3 + 9 + (9 + 8) + (9 + 2\cdot 8) + ... + (9 + (n-2)\cdot 8)$.\\
We denote it as: 
$a_n = -3 + \sum_{m=1}^{n-1}b_m$, where $b_m = 9 + (m-1)8$. So, 
\begin{equation} a_n =  -3 + \sum_{m=1}^{n-1}b_m = -3 + \sum_{m=1}^{n-1}[9 + (m-1)8] = 4n^2 - 3n - 4 \end{equation}
\begin{equation}
\text{d}: \quad -\sum_{n=1}^{\infty} n (4n^2 -3n -4) (-\mbkg2)^{n-1} = -\mf{3((\mbkg2)^2 -8\mbkg2 -1)}{(1 + \mbkg2)^4}   \,.
\end{equation}

\paragraph{Series b.}The last one is b:
\begin{equation}
\begin{split}
\text{b}: \quad &1(-6 \mbkg2) -2(-\mtf{9}{2}(\mbkg2)^2) +3(-1(\mbkg2)^3) -4(\mtf{9}{2} (\mbkg2)^4) +5(12(\mbkg2)^5) -6(\mtf{43}{2}(\mbkg2)^6\\ &+7(33(\mbkg2)^7) -8(\mtf{93}{2}(\mbkg2)^8) = -\sum_{n=1}^{\infty} n a_n  (-\mbkg2)^{n}  \,,
\end{split}
\end{equation}
where:\\
$a_1 = -6$,\\
$a_2 = -6 + \mf{3}{2} = -\mf{9}{2}$,\\
$a_3 = -6 + \mf{3}{2} + (\mf{3}{2} + 2) = -1$,\\
$a_4 = -6 + \mf{3}{2} + (\mf{3}{2} + 2) + (\mf{3}{2} + 2 \cdot 2) = \mf{9}{2}$,\\
$a_5 = -6 + \mf{3}{2} + (\mf{3}{2} + 2) + (\mf{3}{2} + 2 \cdot 2) + (\mf{3}{2} + 3 \cdot 2) = 12 $,\\
...\\
$a_n = -6 + \mf{3}{2} + (\mf{3}{2} + 2) + (\mf{3}{2} + 2 \cdot 2) + ... + (\mf{3}{2} + (n-2) \cdot 2) $.\\
We denote it as: $a_n = -6 + \sum_{m=1}^{n-1}b_m$, where $b_m = \mf{3}{2} + (m-1)2$. So, 
\begin{equation}
a_n = -6 + \sum_{m=1}^{n-1}b_m =   -6 + \sum_{m=1}^{n-1}\ml[ \mf{3}{2} + (m-1)2 \mr]  = \mf{2n^2 -3n -11}{2}
\end{equation}
\begin{equation}
\text{b}: \quad -\sum_{n=1}^{\infty} \mf{n(2n^2 -3n -11)}{2} (-\mbkg2)^{n} = -\mf{3\mbkg2((\mbkg2)^2 +5\mbkg2 +2)}{(1 + \mbkg2)^4}   \,.
\end{equation}

\paragraph{Final formula for $\mk4^{(1,B,\text{NLO})}$.}
We plug in all these results into Eq. (\ref{eq:k4nlo-sers}). Using Eq. \eqref{eq:kappa-bar}, we obtain Eq. \eqref{eq:k4-nlo}.

\section{Limits on $\alpha_k$} \label{appendix:limits-alpha}
In this Appendix, we discuss the limits on the values of the short-range correlation coefficients, $\alpha_k$.

\subsection{Probability distribution}
Firstly, we focus on the discrete probability distribution itself. We straightforwardly differentiate the probability generating function, $H(z) =  \sum_{n=0}^{\infty} P(n) z^n$. We use the facts that $H(z) = e^{G(z)}$ and $G(z) = \sum_{k=1}^{\infty} \mf{(z-1)^k}{k!} \hat{C}_k$, where $G(z)$ is the factorial cumulant generating function. Therefore, the multiplicity probability distribution is given by
\begin{equation}
P(m) = \mf{1}{m!} \ml. \mf{d^m}{dz^m}\ml[ \exp\ml( \sum_{k=1}^{\infty} \mf{(z-1)^k}{k!} \hat{C}_k \mr) \mr] \mr|_{z=0}\,.
\end{equation}
In our case $\hat{C}_k$ is given by Eq. (\ref{eq:fact-cum-glob}). Clearly, $P(m)$ must satisfy the condition $0 \leq P(m) \leq 1$ for all $m$. This is the crucial test for the validity of the set of values of $\alpha_k$'s.\footnote{In practice, we assume that $\alpha_k \neq 0$ for finite $k$, e.g., $k \leq 6$.}

\subsection{Central moments}
The $k$th central moment is defined as $\mu_k = \ml\mla (x - \mla x \mra)^k \mr \mra$.\footnote{It is straightforward to check that: $\mu_2 = \kappa_2$, $\mu_3 = \kappa_3$, $\mu_4 = \kappa_4 + 3\kappa_2^2$, $\mu_5 = \kappa_5 + 10 \kappa_3 \kappa_2$, and $\mu_6 = \mk6 + 15\mk4 \mk2 + 10\mk3^2 + 15\mk2^3$.} Obviously, the even central moments have to be greater than or equal to 0.

First of all, the variance, $\mu_2 = \kappa_2 = \sigma^2 =  \ml\mla (n - \mla n \mra)^2 \mr \mra \geq 0$. Using the definition of the factorial cumulants and assuming the short-range correlations (\ref{eq:fact-cum-glob}), we obtain:
\begin{equation} \label{eq:fact-cum-2}
\hat{C}_2 = \ml. \mf{d^2 G(z)}{dz^2}\mr|_{z=1} = - \mla n \mra^2 + \mla n(n-1) \mra = \alpha_2 \mla n \mra\,.
\end{equation} 
Therefore,
\begin{equation} \label{eq:n2-origin}
\mla n \mra (\alpha_2 + 1) = \mla n^2 \mra - \mla n \mra^2 = \sigma^2 \geq 0 \,.
\end{equation}
This puts the lower limit on $\alpha_2$:
\begin{equation} \label{eq:alpha2-lim}
\alpha_2 \geq -1 \,.
\end{equation}
A similar discussion applies to the 4th and 6th central moments resulting in more complicated relations between $\alpha_k$'s and $\mla n \mra$.

\subsection{Kurtosis-skewness inequality}
There exists an inequality between kurtosis, $K$, and skewness, $S$ \cite{pearson}:\footnote{This inequality can be justified quite easily. Here we follow Ref. \cite{oliveira}. Suppose $x$ is a random variable from the distribution with mean $\mla x \mra$ and standard deviation $\sigma$. Let $y = \mf{x - \mla x \mra}{\sigma}$. Clearly, $\mla y \mra = 0$, $\sigma^2_y = \mla y^2 \mra = 1$. We use the Cauchy-Schwartz inequality for probability theory, $\mla a b \mra^2 \leq \mla a^2 \mra \mla b^2 \mra$, where $a$ and $b$ are the random variables. Let $a = y$, $b = y^2 - 1$. This brings us to $\mla y^4 \mra \geq \mla y^3 \mra ^2 + 1\,$, being equivalent to (\ref{eq:ineq-centr-mom}). }
\begin{equation} \label{eq:ineq}
K \geq S^2 + 1\,,
\end{equation}
or in terms of the central moments:
\begin{equation} \label{eq:ineq-centr-mom}
\mf{\mu_4}{\mu_2^{2}} \geq \mf{\mu_3^2}{\mu_2^3} + 1\,,
\end{equation}
or in terms of cumulants:
\begin{equation} 
\mf{\kappa_4}{\kappa_2^{2}} + 2 \geq \mf{\kappa_3^2}{\kappa_2^3}\,.
\end{equation}
This condition also gives nontrivial relations between $\alpha_k$'s.

\bibliography{baryon_short_correction_v7_ref}


\end{document}